# Current driven discontinuous insulator-metal transition and colossal low-field magnetoresistance in $Sm_{0.6}Sr_{0.4}MnO_3$

A. Rebello and R. Mahendiran [*]

Department of Physics and NUS Nanoscience and Nanotechnology Initiative (NUSNNI), National University of Singapore, 2 Science Drive 3, Singapore - 117542, Singapore

**Abstract**

It is shown that with increasing magnitude of current ($I$), resistivity of $Sm_{0.6}Sr_{0.4}MnO_3$ transforms from a smooth to a discontinuous insulator-metal transition which is also accompanied by an abrupt decrease in temperature of the sample. We report colossal low-field magnetoresistance under a high current bias (-99% at $H = 0.5$ T and 70 K) and electoresistance (-8000 % at $H = 0$ T and 60 K) for $I = 11$ mA. We interpret our observations in terms of current induced supercooling of the high temperature paramagnetic phase and enlargement of volume fraction of the ferromagnetic phase under a magnetic field.

---

[*] Corresponding author – phyrm@nus.edu.sg

The dramatic decrease of electric resistivity above a threshold dc electric field reported in colossal magnetoresistive oxide $Pr_{0.7}Ca_{0.3}MnO_3$ spurred a lot of interest in manganites owing to its technological importance in non volatile resistive memory devices and fundamental physics involved. [1,2] Although intensive experimental and theoretical investigations in the past one decade have helped us to identify mechanisms of colossal magnetoreresistance,[3] the origin of electroresistance still remains elusive. The electric-field induced insulator to metal transition observed in the charge ordered insulating manganite $Pr_{0.7}Ca_{0.3}MnO_3$ was initially attributed to electric field induced melting of charge ordering similar to the effect of magnetic field,[1] other explanations such as formation of filamentary conducting channels,[4] electric field induced depinning of charge density waves,[5] $e_g$-orbital reorientation,[6] have been proposed.

A large decrease of the four probe resistance under current bias was also reported in manganites such as $La_{1-x}A_xMnO_3$ (A= Sr, Ca, x ≈ 0.2-0.33) which are a ferromagnetic metals at low temperatures.[7,8,9] A naïve explanation is that forced motion of spin polarized $e_g$-electron under the influence of high current enhances the $e_g$-band width, reduces scattering and hence the resistivity. Though few researchers suggested that self Joule heating is important and it could lead to current- induced negative electroresistance in manganites,[10,11,12,13,14] majority of researchers ignore the relevance of the Joule heating. Hence, there is an urgent need to identify various key factors that influence electroresistance that will lead to better understanding of the electroresistance phenononenon in these materials.

In this report, we investigate the effect of dc current on the temperature dependence of the resistivity in $Sm_{0.6}Sr_{0.4}MnO_3$. The $Sm_{1-x}Sr_xMnO_3$ (x = 0.35-0.5) series is quite interesting because there are evidences for presence of short range charge ordering[15] as well as ferromagnetic clusters[16] above the ferromagnetic transition temperature A field-induced metamagnetic transition in the paramagnetic state resulting from the collapse of short range charge ordering and increase in the size of ferromagnetic clusters leads to a large magnetocaloric effect in x = 0.4 and 0.5.[17] However, there is no earlier report on the effect of dc current on electrical resistivity in this series and it is the motivation of the present work.

We have simultaneously measured the four probe resistivity of rectangular polycrystalline sample of $Sm_{0.6}Sr_{0.4}MnO_3$ of dimensions 3 x 3 x 10 mm$^3$ and temperature of the sample by thermally anchoring a Pt-100 temperature sensor of size 2 x2 x3 mm$^3$ on the top of the sample with a thermal conductive GE-7031 grease. The temperature of the sample measured by the Pt-sensor is denoted as $T_S$ and the temperature recorded by a cernox-sensor in the commercial superconducting magnet (Physical Property Measuring System, Quantum Design, USA) is denoted as the base temperature T throughout this manuscript.

Figure 1 shows the temperature dependence of the dc resistivity, $\rho(T)$, for different values of the dc current ($I$ = 0.1 mA to 13 mA). The $\rho(T)$ for both $I$ = 0.1 mA and 1 mA closely overlap on each other and show a smooth semiconductor-metal transition with a peak in $\rho(T)$ occurring around $T_{IM}$ = 115 K while cooling. The I-M transitions at $I$ = 0.1 and 1 mA are weakly hysteretic while cooling and warming. However, the I-M transition shifts to lower temperature and the transition to the metallic state becomes discontinuous with a wider hysteresis behavior upon increasing amplitude of the current. For example, the $\rho(T)$ for $I$ = 13 mA upon cooling from 200 K shows a discontinuous decrease at $T_{IM}$ (cooling) 40 K and then continues to decrease gradually with further lowering temperature. Upon warming, a reverse transition from the metallic to the insulating state takes place discontinuously around $T_{IM}$ =75 K. While $\rho(T)$ for $I > 5$ mA is lower than for I < 1 mA below 160 K, the peak value of the resistivity just before the discontinuous jump is not severely affected by the amplitude of the current except for $I = 13$ mA. Surprisingly, the discontinuous I-M transition is also accompanied by a discontinuous decrease in the temperature of the sample ($T_S$) as shown in Fig. 1(b) where the actual temperature of the sample ($T_S$) is plotted against the base temperature of the sample puck ($T$) which is generally measured in the Physical Property Measuring System. It can be noted that $T_S >> T$ before the discontinuous decrease takes place. As the amplitude of the current decreases the discontinuity in the temperature also decreases and finally both $T_S$ and $T$ closely match for $I = 0.1$ mA and 1 mA. In Fig. 3 (a) we plot the $T_{IM}$ while cooling and warming as a function of current. Fig 3(b) shows the width of the hysteresis $\Delta T = T_C - T_W$, (where $T_C$ and $T_W$ are the base temperatures at which discontinuous change in the resistivity takes place while cooling and warming, respectively) plotted against the amplitude of the current. We see that the width of the hysteresis increases with the amplitude of the current.

In Fig. 2(a) and 2(b), we compare the effect of dc magnetic field ($H$) on $\rho(T)$ for $I = 0.1$ mA and $I = 11$ mA, respectively. The insets show $T_S$ versus $T$ in both cases. The application of H dramatically decreases the peak value of the resistivity for $I = 0.1$ mA and the position of the peak shift to high temperature with increasing value of $H$ which is clearly visible for $H > 2$ T. The effect of $H$ is more dramatic when $I = 11$ mA. While the discontinuous I-M transition while cooling occurs at 52 K for $H = 0$ T, a small magnetic field of $H = 0.5$ T dramatically shifts the I-M transition to 75 K. As $H$ increases further, the discontinuous I-M transition shifts upward in temperature and the width of the hysteresis decreases. Finally, the discontinuous I-M transition transforms into a smooth I-M transition for $H \geq 3$ T. Fig. 2(c) compares the magnetic field dependence of the $T_{IM}$ and $T_{MI}$ while cooling and warming .for both the values of currents. The transition temperatures both cooling and warming for $I$ = 1 mA are higher than for $I$ = 13 mA. While the transition temperature increases nearly linearly for $I$ = 1 mA, the $T_{IM}$ while cooling shifts more rapidly than $I_{MI}$ while warming for $I$ = 13 mA and for $H$ < 2 T. We also note that the curves for $I$ =1 mA and 13 mA have nearly same slope for $H$ > 2 T. Fig. 2(d) shows the difference $\Delta T$ = $T_C$ - $T_W$. The $\Delta T$ for $I$ = 11 mA decreases rapidly with the magnetic field for $H$ < 2 T.

Fig. 4(a) shows the percentage change in the electroresistance (ER (%) = [ρ(I mA)–ρ(0.1 mA)]/ρ(0.1 mA)x100) as a function of temperature in absence of magnetic field. We show the data only for cooling. While the main panel shows ER from 10 to 200 K, the inset shows the behavior of ER expanded between 100 K and 200 K. As the inset suggests, the ER is positive and it increases below 180 K and goes through a maximum around 120 K. The magnitude of the ER increases with the amplitude of the current but the position of the maximum is independent of the current. A maximum ER of 75 % at 120 K is obtained for $I$ = 13 mA. However, below 100 K (see the main panel), the ER becomes negative and exhibits a sharp peak before going to zero. The position of the negative peak shifts to lower temperature and it's the peak value decreases with increasing amplitude the current. A maximum ER of – 13000 % occurs at 37 K for $I$ = 13 mA and -8000 % around 67 K for $I$ = 11 mA.

Fig. 4(b) shows the magnetoresistance (MR(%) = [ρ(H)–ρ(0)]/ρ(0)x100) where ρ(H) and ρ(0) are the resistivities in a magnetic field of $H$ Tesla and $H$ = 0 Tesla, respectively for $I$ = 0.1 mA. The MR at $H$ = 0.5 T is negligible at high temperatures but it increases below 125 K and exhibits a peak around 100 K upon cooling. While the position of the peak is not very much affected with increasing $H$, the magnitude of the peak increases. A maximum MR of 90 % occurs around $T = 110$ K for $H = 5$ T. The MR at the lowest temperature ($T = 10$ K) and at a high temperature (T= 200 K) also increase with H. In Fig. 3(c) we show the MR for $I = 11$ mA. Unlike a clear peak for $I = 0.1$ mA, the MR at $H = 0.5$ T increases discontinuously around 75 K, shows a plateau between 75 K and 55 K before showing a discontinuous decrease. The most spectacular observation is that the MR reaches nearly 100 % in the plateau region even for $H = 0.5$ T. As H increases, the plateau region extends in temperature on the high temperature side because the temperature at which MR increases abruptly also increases with the magnetic field. However, on the low temperature side, the temperature at which $\rho(T)$ decreases discontinuously is not affected by the value of H. The extension of the temperature range over which the MR can remain high, i.e. nearly 100 % is very interesting for practical applications.

The two central questions regarding all the above observations are (1) what the role of joule heating is and (2) why ρ(T) becomes discontinuous with increasing strength of the dc current ? Cooling the sample from 300 K in presence of a large current (for example, $I = 13$ mA ), streamlines electron hopping in the direction of the electric field and also leads to Joule heating as the temperature decreases. Because $d\rho/dT < 0$ in the paramagnetic state Joule heating leads to decrease in the resistivity. While Joule heating is insignificant above $T \approx 180$ K, Joule power dissipated [P(T) = $I^2R(T)$] increases for $T < 180$ K leading to nucleation of "hot electrothermal domains" which have low resistivities than the rest of the matrix whose average temperature is low ("cold matrix").[18] The preferential flow of current through the low resistive hot electrothermal domains leads to lowering of the resistivity as suggested by the ρ(T) curves for $I > 1$ mA which lie below the curve for $I = 1$ mA for $T <$ 180 K.  The size of the electrothermal domains also increase with lowering temperature. Although whole volume of the sample becomes ferromagnetic and metallic below $T_C$ = 115 K for $I$ < 1mA, annealing under a large dc current leads to supercooling of the hot electrothermal domains while the "cold matrix" become ferromagnetic below $T_C$. Since electrothermal domains are paramagnetic, both paramagnetic and ferromagnetic phases coexist below $T_C$. Continuous cooling below $T_C$ leads to the collapse of the hot electrothermal domains at a specific temperature which leads to a discontinuous I-M transition. The sample can be considered as homogeneous ferromagnet for temperature below the discontinuous I-M transition. The abrupt decrease of the sample's temperature suggests that electrothermal domains are probably macroscopic  in size and occupy a large fraction of the volume below $T_C$ rather than filamentary in nature.  The collapse of hot electrothermal domains can lead to a sudden increase in volume fraction of the ferromagnetic phase which has lower unit cell volume than the paramagnetic phase.[19] This structural changes can aid to the sharpness of the transition.

While warming from low temperature, the paramagnetic electrothermal domains nucleate and grows in the ferromagnetic matrix. When the paramagnetic domains engulfs the ferromagnetic phase completely, the resistance increases abruptly. This reverse transition will be accompanied by an abrupt increase in the volume of the sample.  L. Sudeendra and C. N. R. Rao [20] also observed current driven first-order I-M transition in $La_{0.67}Ca_{0.23}MnO_3$ single crystal under a large current of 60 mA and attributed the electroresistance to softening of spin wave excitation spectrum which would enhance spin fluctuation and lower the $T_C$. The reported results can be also interpreted as a consequence Joule heating effect and current induced phase separation. Cooling under a magnetic field not aligns the domain magnetization of the ferromagnetic phase but also tends to align $t_{2g}^3$ spins and enhances hopping of $e_g$-electron in the hot eletrothermal domains. As a result, Joule heating in the electrothermal decreases and the collapse of hot electrothermal domains occur at high temperature. Hence, the temperature corresponding to the discontinuous I-M transition shifts quickly to high temperature with low cooling fields ($H < 1$ T). At higher magnetic fields, hot electrothermal domains no longer exist and the sample is a homogeneous ferromagnet at low temperature. Hence, the discontinuous I-M transition becomes a continuous I-M transition.

In summary we have shown that the current driven discontinuous insulator-metal transition and colossal electroresistance in $Sm_{0.6}Sr_{0.4}MnO_3$ is primarily caused by the Joule heating which leads to coexistence of ferromagnetic metallic and paramagnetic semiconducting phases over a certain temperature range. Our results indicate that the input power to the sample, and hence the Joule heating can be fine tuned to enhance the low-field magnetoresistance. The present work also stresses the importance of measuring the actual temperature of the sample directly instead of relying on the temperature recorded by a commercial instrument in elucidating the origin of the colossal electroresistance and non linear electrical transport in manganites.

This work was supported by the office of the DPRT (NUS) through the grant, NUS-YIA-. R144-000-197-123.

**Figure captions**

Fig. 1 (a) Temperature dependence of the resistivity, ρ(T) under different dc current ($I$) in the absence of an external magnetic field and (b) sample temperature ($T_S$) recorded by a Pt- temperature sensor attached directly on the top surface. The $T$ on the y-axis represents the temperature recorded by the commercial cryostat used in this work.

Fig. 2 (a) Current dependence of the temperatures corresponding to insulator-metal transition upon cooling ($T_C = T_{IM}$) and the reverse transition ($T_W = T_{MI}$) upon warming. (b) The difference $\Delta T = T_C - T_W$ as a function of current (c) Magnetic field dependence of $T_{IM}$ and $T_{MI}$ for $I$ = 1 and 11 mA. (d) ΔT as a function of H.

Fig. 3 Temperature dependence of the resistivity, ρ(T), for (a) $I$ = 1 mA and (b) $I$ = 11 mA under external magnetic fields. The insets show temperature $T_S$ of the sample under different magnetic fields.

Fig. 4 Temperature dependences of the (a) Electoresistance (ER) for different strengths of dc current ($I$) in absence of a magnetic field. The inset shows the ER above 100 K in an expanded scale. Note that while the ER in the main panel is expressed as $10^3$ % it is written in % for the inset. (b) Magnetoresistance (MR) for $I$ = 1 mA and (c) 11 mA.

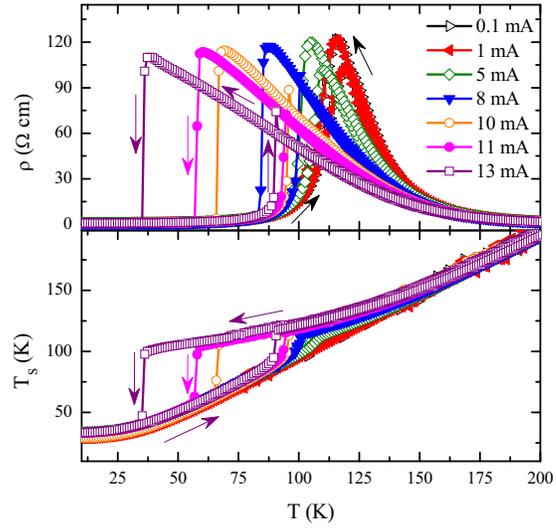

Figure 1

A. Rebello *et al*.

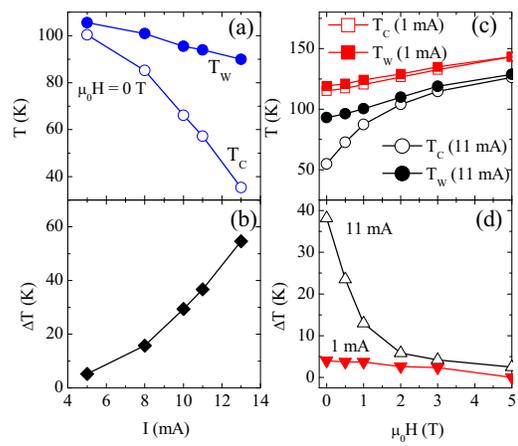

Figure 2

A. Rebello et al.

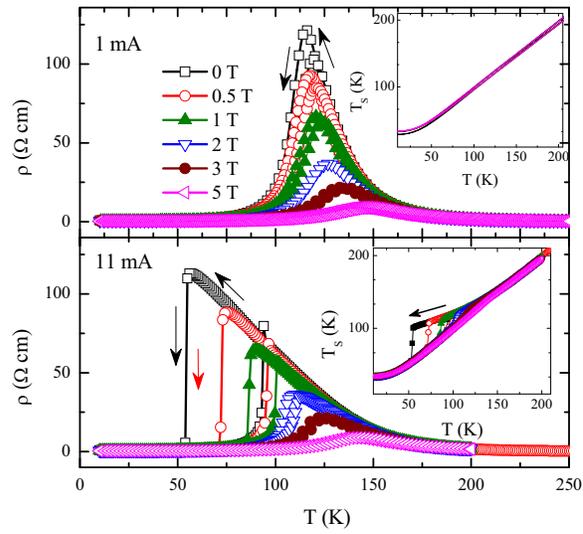

Figure 3

A. Rebello *et al.*

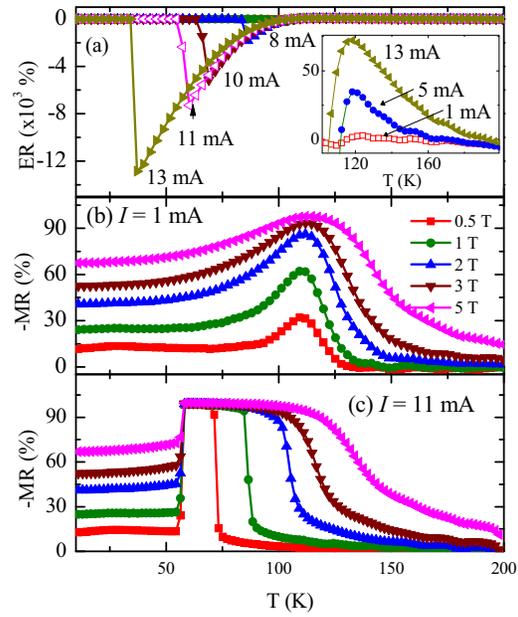

Figure 4

A. Rebello *et al*.